\documentclass[a4paper,nodate]{sab} 

\usepackage{graphicx} 
\usepackage{txfonts,natbib} 
\bibpunct{(}{)}{;}{a}{}{,} 
\usepackage{hyperref}
\usepackage{url}
\usepackage[T1]{fontenc} 
\usepackage{t1enc} 
\usepackage{xcolor}
\usepackage{ulem}
\renewcommand\ackname{Acknowledgements.} 
\renewcommand\andname{\&}
\renewcommand\keywordname{Keywords.} 
\renewcommand\figureptname{Figure}
\renewcommand\tableptname{Table}

\renewcommand\noteaddname{Note added to proofs.}
\renewcommand\andname{\&}

\idline{Boletim da Sociedade Astron\^omica Brasileira, {\bf 33}, no. 1, 168-172}{1}
\begin{document} 
\title{What about high redshift sources in the Main Sequence of quasars?} 

\author{A. Deconto-Machado \inst{1} \and A. del Olmo Orozco \inst{1} \and P. Marziani \inst{2}} 
\institute{Instituto de Astrofísica de Andalucía, IAA-CSIC, Glorieta de la Astronomia, s/n 18008, Granada Spain \hspace{35mm}\email{adeconto@iaa.es; chony@iaa.es}  \and National Institute for Astrophysics (INAF), Osservatorio Astronomico di Padova, vicolo dell' Osservatorio, IT 35122, Padova, Italy \email{paola.marziani@inaf.it}}
\date{Received } 

\Abstract {Much effort has been done in order to better understand the active galactic nuclei mechanisms behind the relativistic jets observed in radio-loud sources. These phenomena are commonly seen in luminous objects with intermediate/high redshift such as quasars, so that the analysis of the spectroscopic properties of these sources may be a way to clarify this issue. Measurements are presented and contextualized taking advantage of the set of correlations associated with the quasar Main Sequence (MS), a parameter space that allows to connect observed properties to the relative relevance of radiative and gravitational forces. In the redshift range we consider, the low-ionization HI Balmer line H$\beta$ is shifted into the near infrared. Here we present  first results of a sample of 22 high-luminosity quasars with redshift between 1.4 and 3.8. Observations covering the H$\beta$ spectral region were collected with the IR spectrometer ISAAC at ESO-VLT. Additional data were collected from SDSS in order to cover the UV region of our sources. The comparison between the strong C \textsc{IV} $\lambda$1549 high-ionization line and H$\beta$ in terms of line widths and shifts with respect to the rest-frame leads to an evaluation of the role of radiative forces in driving an accretion disk wind. While for non-jetted quasars the wind properties have been extensively characterized as a function of luminosity and other physical parameters, the situation is by far less clear for jetted sources. The overarching issue is the effect of the relativistic jets on the wind, and on the structure of the emitting region in general. We present results from our analysis of the optical and UV line profiles aimed to identifying the wind contribution to the line emission.}{Muito esforço vem sendo realizado para entender os mecanismos que envolvem os núcleos ativos de galáxias que estão por trás dos jatos relativisticos observados em objetos \textit{radio-loud}. Esses fenômenos são normalmente detectados em galáxias luminosas de intermediário/alto redshift como os quasares, por exemplo. Uma maneira de clarificar esse comportamento é através da análise das propriedades espectroscópicas desses objetos. As medidas aqui apresentadas são contextualizadas com as correlações associadas à \textit{Main Sequence} (MS) de quasares, um espaço de parâmetros que nos permite conectar propriedades observacionais com a relativa relevância das forças radiativas e gravitacionais. No intervalo de \textit{redshift} que consideramos, a linha de Balmer de baixa ionização H$\beta$ apresenta um \textit{shift} para o infravermelho próximo. Neste trabalho, apresentamos os resultados iniciais de uma amostra de 22 quasares de alta luminosidade com \textit{redshift} entre 1.4 e 3.8. As observações que compreendem o intervalo ótico foram coletadas a partir do espectrógrafo em infravermelho ISAAC do ESO-VLT. Dados adicionais para cubrir o intervalo em UV da amostra foram obtidos através do SDSS. A comparação entre a linha de alta ionização C \textsc{IV} $\lambda$1549 e H$\beta$ em termos de largura e \textit{shifts} das linhas com respeito ao \textit{rest-frame} leva a uma avaliação do papel das forças radiativas no processo de produção dos ventos decorrentes do disco de acreção. Enquanto que para quasares que não apresentam jatos as propriedades dos ventos foram extensivamente caracterizadas em função da luminosidade do objeto e outros parâmetros físicos, a situação para os quasares que apresentam jatos é muito menos conhecida. A principal questão é o efeito dos jatos relativísticos nos ventos galácticos, e a estrutura da região emissora em geral. Apresentamos resultados da nossa análise nas regiões espectrais que abrangem o ótico e o ultravioleta com o intuito de identificar a contribuição de \textit{outflows} nas linhas de emissão.}

\keywords{Quasars: general -- Quasars: emission lines -- Galaxies: active}


\maketitle 

\section{Introduction}

\begin{figure*}[h!]
    \centering
    \includegraphics[width=0.985\linewidth]{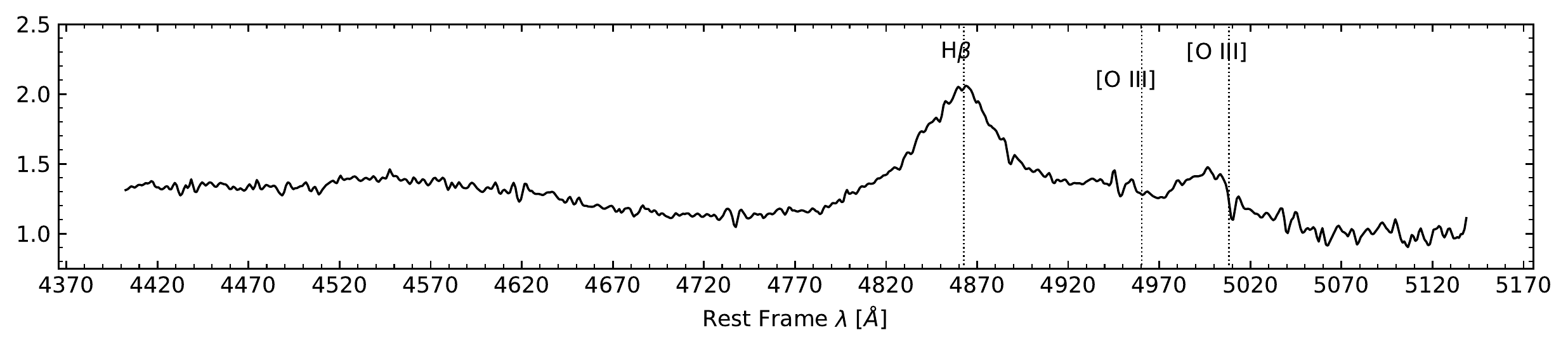}
    \includegraphics[width=0.985\linewidth]{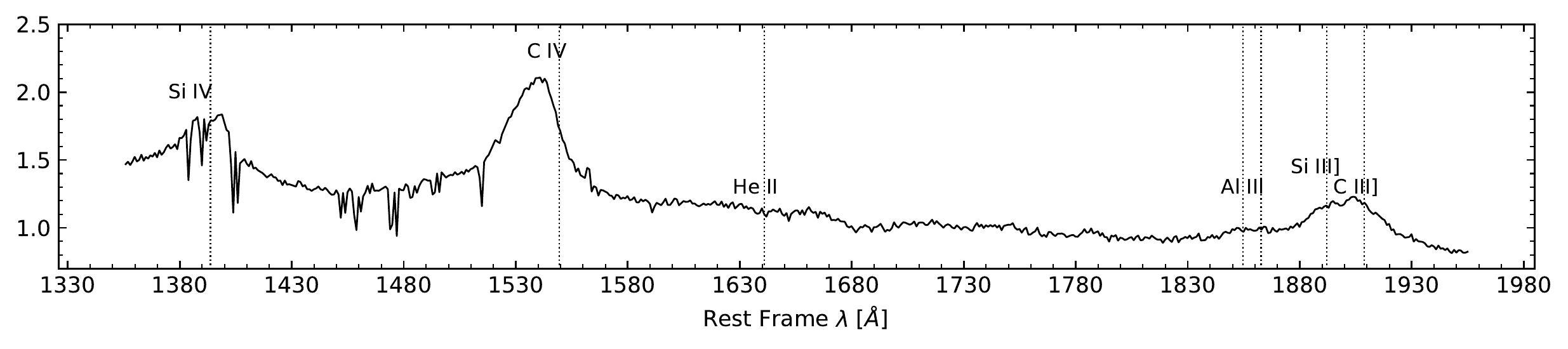}
    \caption{Optical and UV spectra (top and bottom panels, respectively) of SDSSJ161458.33+144836.9 as an example of our sources. Vertical dotted lines indicate the rest-frame of the main emission lines.}
    \label{fig:spectra}
\end{figure*}

\par The fourth-dimensional Eigenvector 1 (4DE1) has been proposed by \cite{sulentic_2000} and allows for the classification of the different low-redshift quasars.
The 4DE1 consists of a correlation space that takes into account several key observational   optical, UV, and X-ray measures that are related to outflow prominence, accretion mode and other physical parameters.
\par One important measurement in the 4DE1 context is the full width at half maximum (FWHM) of H$\beta$, which can be interpreted as a measure of virialized motions in the accretion disc \citep{marziani_2018}. In addition, another optical property of the Eigenvector 1 is the ratio between the intensities of the Fe \textsc{II} blend at 4570\r{A} and H$\beta$ (R$_{\rm{Fe II}}=I(\rm{Fe \textsc{II}}\lambda4570)$/$I(\rm{H}\beta)$). This property can reveal more details of the Broad Line Region (BLR), as it is dependent on the ionization state, the electron density, and column density of this region.
\par The optical plane of the 4DE1 which considers the FWHM(H$\beta$) vs. R$_{\rm{Fe \textsc{II}}}$ corresponds to the Main Sequence (MS) of quasars and can be used to classify these objects into Population A and Population B according to their significant spectroscopic differences  \citep{sulentic_2000, Zamfir_2008}. Pop. B quasars tend to present lower Fe II contribution and are more disk-dominated when compared to Pop. A objects \citep{sulentic_2011}. At low-redshift, the strongest outflows are seen in Pop. A, especially in extreme Pop. A \citep{martinezaldama_2018}. Low-redshift Pop. B quasars usually do not present significant blue components that are representative of outflowing gas in the spectra \citep{sulentic_2017}.
\par Pop. A quasars present low FWHM of H$\beta$\ (usually lower than 4000 km/s).  Spectral types are defined by increasing R$_{\rm{Fe \textsc{II}}}$, going from A1 to A4. On the other hand, Pop. B sources are the ones that present a very wide range of FWHM(H$\beta$), always larger than 4000 km/s and spectral types are defined in terms of increasing FWHM(H$\beta$), going from B1 to B1++. Also, strong wind components are expected in Pop. A and Pop. B in a high-redshift and high-luminosity context \citep{vietri_2018, sulentic_2017}. In these situations there is a broadening of the emission line components such as H$\beta$ and C IV$\lambda$1549 \citep{sulentic_2007}. In the present work, we report on the behaviour of a sample of 22 high-redshift sources in the context of the Main Sequence of quasars.

\section{Sample and Observations}

\par The sample consists of 22 QSOs with a redshift range of $z=1.4-3.8$ and includes both radio-loud and radio-quiet sources. The infrared observations were performed at one of the eight-meter VLT telescopes under the programmes 083.B-0273(A) and 085.B-0162(A) from the European Southern Observatory (ESO). The spectroscopic reduction of the new VLT observations were performed in the standard way using the routines of the astronomical package IRAF. UV spectra were selected mainly from the archive of the Sloan Digital Sky Survey (SDSS). The UV spectral range were observed in the optical domain due to the high redshift of the sample and includes emission lines such as C IV $\lambda1549$, He II $\lambda1640$, and Si IV+O IV]$\lambda1500$.

\par We have also performed the redshift ($z$) estimation based on the H$\beta$ emission line and the same value of $z$ was applied for the optical and UV spectra. One example of optical and UV spectra at rest-frame is shown in Fig. \ref{fig:spectra}. We present the data for SDSSJ161458.33+144836.9, which has a redshift $z=2.5698$. Regarding the radio frequencies data, they were collected from the Faint Images of the Radio Sky at Twenty-Centimeters (FIRST, \cite{becker_1994}) and The NRAO VLA Sky Survey (NVSS, \cite{condon_1998}) survey archives.

\section{Spectral analysis}

\par In order to perform the spectral analysis, we implemented a minimum-$\chi^2$ fit of the continuum and of the individual spectral line components through the \texttt{specfit} routine from IRAF. By performing these fittings, we are able to evaluate the FWHM, peak wavelength, and flux intensities of all the line components. Here we will discuss the fittings of H$\beta$+[OIII]\-$\lambda\lambda4959,5007$ and C IV$\lambda1549$+He II$\lambda$1640 regions.
\par For the optical range, we have fit the H$\beta$+[O III]$\lambda\lambda4959,5007$ range. In general, we consider three/four different profiles: a blueshifted component (BLUE), a symmetric and unshifted broad component (BC), a redshifted very broad component (VBC, only for Pop. B sources), and a narrower component (NC), apart from a power law continuum and a scalable Fe II emission template. We used the Fe II template from \cite{borosongreen_1992} updated with the improvements of \cite{marziani_2009}. BLUE components are seen as a good indicator of non-virial motion and/or outflowing gas from the central region and the BCs are thought to account for the virialized motion from the Broad Line Region (BLR).
\par Depending on the classification of the source, we fit the spectra in different ways. In the case the quasar is a Pop. A, the BC is represent by a Lorentzian profile and all the other components by Gaussians. On the other hand, if the source is a Pop. B, then all the components are fitted by Gaussian-like profiles. Our initial guess is that the NC and BC are located in the rest-frame, differently from the BLUE and VBC. Apart from the shift towards lower wavelengths, blueshifted components can also present blueward asymmetries.

\par [O III]$\lambda\lambda$4959,5007 emission lines were fitted following a similar approach of H$\beta$. In this case, the full profile is composed by a narrow component (NC) and a semi-broad component (SBC) for the two populations. The SBC is considered to present a Lorentzian shape for Pop. A and a Gaussian profile for Pop. B sources. We assume that narrow components of H$\beta$ and [OIII]$\lambda\lambda$4959,5007 share the same shift, FWHM and asymmetry. 


\par Regarding the UV spectra, we fit the C IV$\lambda1549$+He II$\lambda1640$ emission lines. In the case the source is Pop. A, we reproduce the full profile by including only two profiles: a Lorentzian-like BC placed at rest-frame and a blueshifted component. For Pop. B, we include a VBC in addition to the BC+BLUE and the BC present a Gaussian shape.

\section{Results and Discussion}

\begin{table}[h!]
    \centering
    \resizebox{\linewidth}{!}{
    \begin{tabular}{|l|c|c|c|}
    \hline
    \textbf{Source identification} & \textbf{Pop.} & \textbf{FWHM(H$\beta_{\rm{full}}$)} & \textbf{FWHM(C IV$_{\rm{full}}$)} \\
    & &\textbf{ [km s$^{-1}$]} & \textbf{[km s$^{-1}$]} \\
    \hline
        HE0001-2340 & B & 6632 $\pm$ 529 & - \\
        \hline
        $[\rm HB89]$ 0029+073 & B & 4971 $\pm$ 373 & -\\
        \hline
        CTQ 0408 & B & 6405 $\pm$ 499 & -\\
        \hline
        SDSSJ005700.18+143737.7 & A & 3830 $\pm$ 342 & 8648 $\pm$ 562\\
        \hline
        H0055-2659 &  B & 5342 $\pm$ 458 & -\\
        \hline
        SDSSJ114358.52+052444.9 &  B & 5944 $\pm$ 463 & 7078 $\pm$ 766\\
        \hline
        SDSSJ115954.33+201921.1 & B & 6944 $\pm$ 543 & 5741 $\pm$ 585\\
        \hline
        SDSSJ120147.90+120630.2 & B & 5839 $\pm$ 521 & 4995 $\pm$ 475\\
        \hline
        SDSSJ132012.33+142037.1 & A & 3450 $\pm$ 314 & 5549 $\pm$ 399\\
        \hline
        SDSSJ135831.78+050522.8 & A & 3548 $\pm$ 312 & 7365 $\pm$ 655\\
        \hline
        Q1410+096 & A & 3394 $\pm$ 299 & 6311 $\pm$ 552\\
        \hline
        SDSSJ141546.24+112943.4 & B &5595 $\pm$ 549 & -\\
        \hline
        B1422+231 & A & 5136 $\pm$ 452 & -\\
        \hline
        SDSSJ153830.55+085517.0 & A & 5107 $\pm$ 450 & 5819 $\pm$ 448\\
        \hline
        SDSSJ161458.33+144836.9 & A & 3722 $\pm$ 327 & 5790 $\pm$ 554\\
        \hline
        PKS1937-101 & A & 5298 $\pm$ 466 & -\\
        \hline
        PKS2000-330 & A & 3138 $\pm$ 276 & 4950 $\pm$ 346\\
        \hline
        SDSSJ210524.47+000407.3 & A & 5032 $\pm$ 443 & -\\
        \hline
        SDSSJ210831.56-063022.5 & A & 5243 $\pm$ 365 & 9389 $\pm$ 519\\
        \hline
        SDSSJ212329.46-005052.9 & A & 5171 $\pm$ 405 & 7978 $\pm$ 524\\
        \hline
        PKS2126-15 & A & 4391 $\pm$ 451 & -\\
        \hline
        SDSSJ235808.54+012507.2 & A & 4098 $\pm$ 360 & 6504 $\pm$ 398\\

    \hline
    \end{tabular}
    }
    \caption{Measurements of the FWHM of full profile of H$\beta$ and C IV$\lambda$1549 emission lines for the complete sample. }
    \label{tab:my_label}
\end{table}

\subsection{The multicomponent fittings}
\par
\begin{figure}
    \centering
    \includegraphics[width=0.495\linewidth]{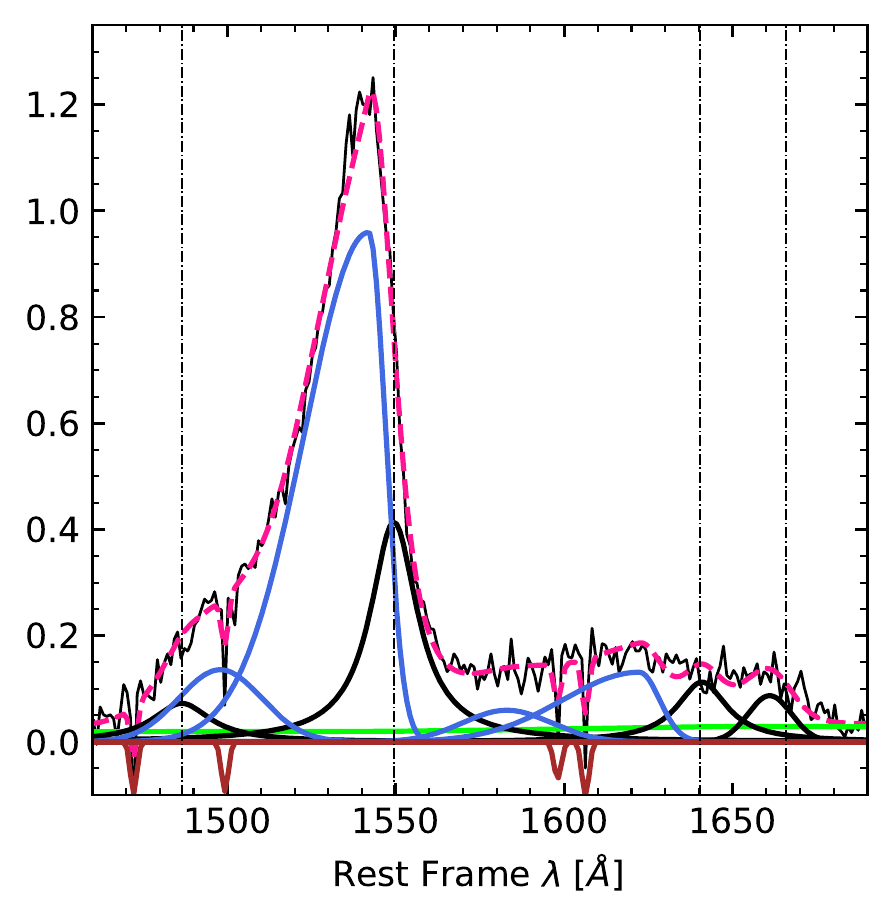}
    \includegraphics[width=0.495\linewidth]{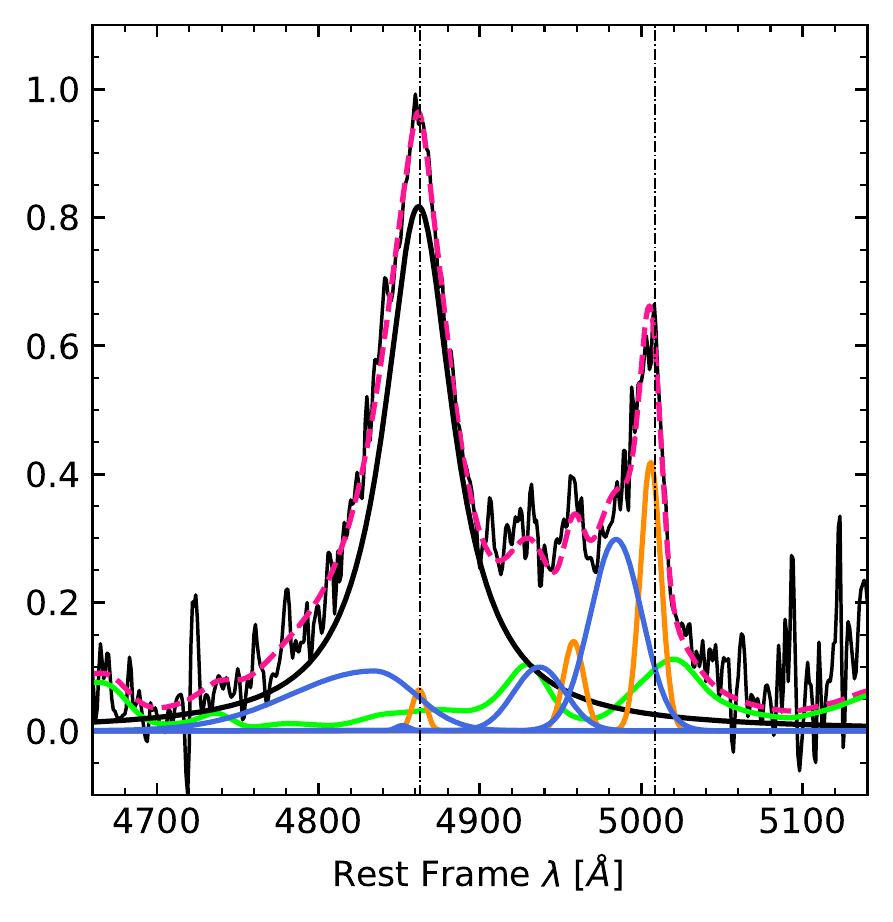}
    
    \includegraphics[width=0.495\linewidth]{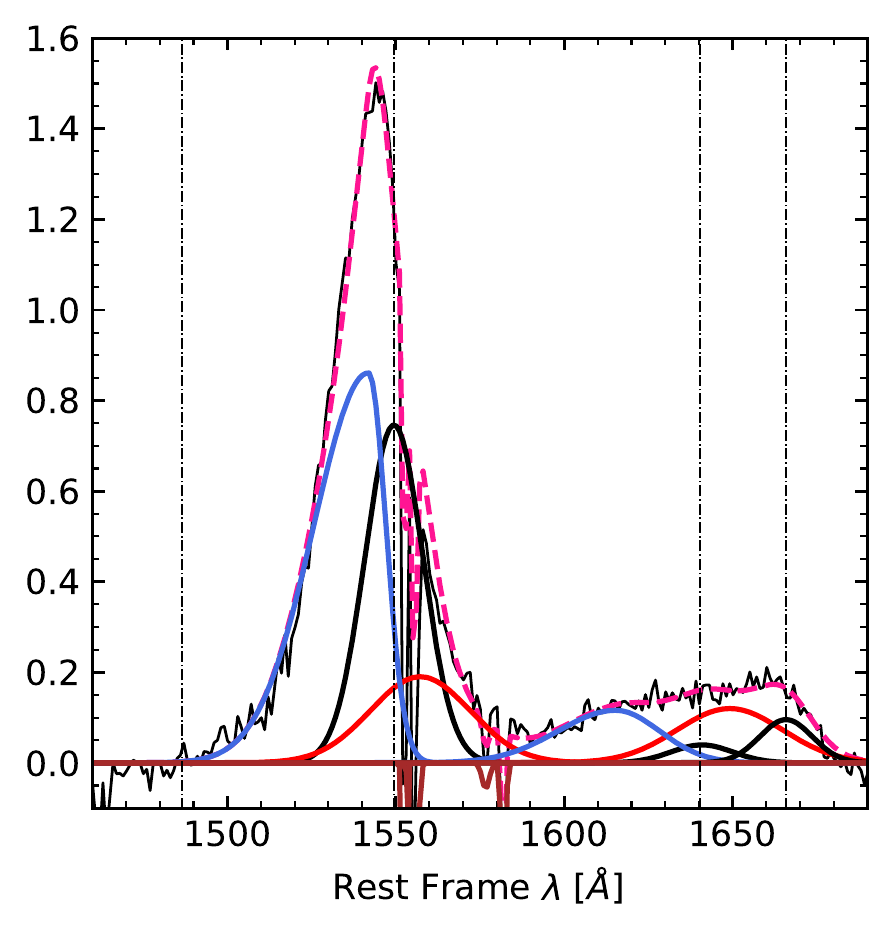}
    \includegraphics[width=0.495\linewidth]{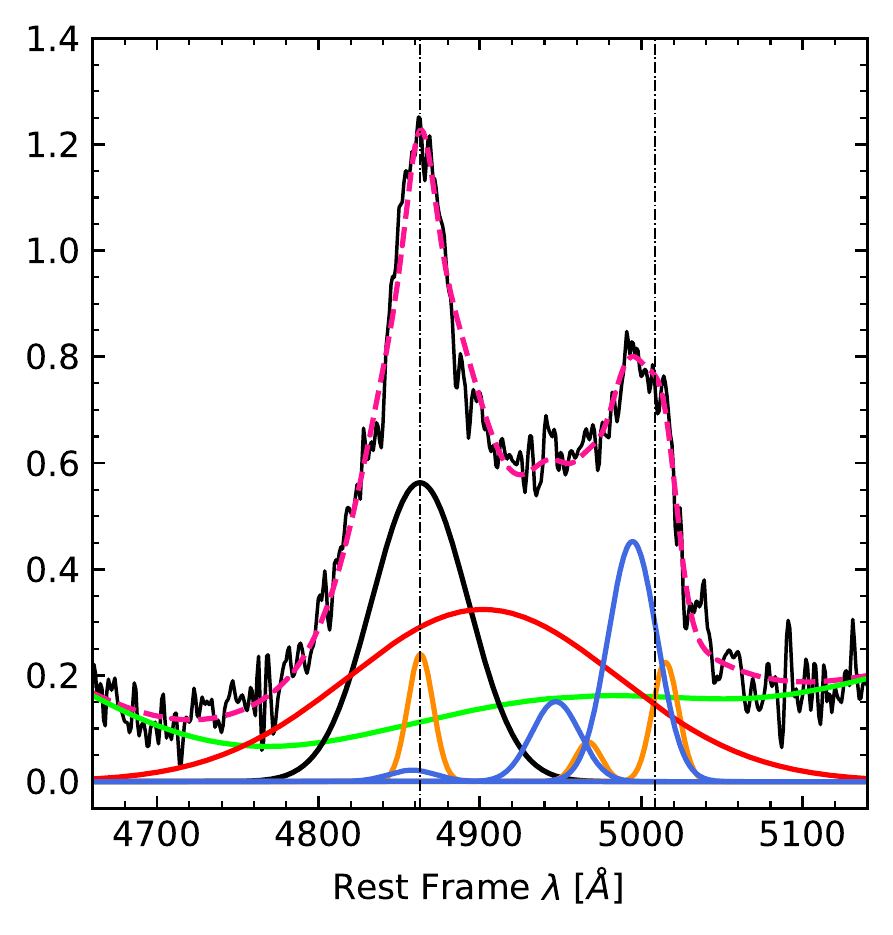}
    
    \caption{\textit{From left to right:} Fittings of the C IV$\lambda1549$+He II$\lambda1640$ and H$\beta$+[O III]$\lambda\lambda4959,5007$ emission lines. \textit{Top panels:} Pop. A source SDSSJ132012.33+142037.1. \textit{Bottom panels:} Pop. B source SDSSJ120147.90+120630.2. Pink dashed lines show the final fitting. Black and blue lines indicate broad and blueshifted components, respectively. Very broad components are shown by red lines. Orange and green lines represent narrow and Fe II components. Residuals are shown in the bottom of each plot.}
    \label{fig:PopAeB}
\end{figure}

 Examples of the fittings performed for the two populations A and B are shown in Fig. \ref{fig:PopAeB}. We choose to present SDSSJ132012.33+142037.1 as an example of Pop. A and SDSSJ120147.90+120630.2 as Pop. B. As it can be seen, in the case of the Pop. A source the H$\beta$ BC profile represents the majority of the contribution to the full emission line. In this source there is also a small contribution of the H$\beta$ blueshifted component, which presents a strong asymmetry towards lower wavelengths. For SDSSJ120147.90+120630.2, which is the Pop. B example, the H$\beta$ full profile is strongly affected by the presence of a redshifted very broad component. This effect is seen in all Pop. B sources from our sample.

\par In the two populations, it is expected that [O III]$\lambda\lambda$4959,5007 present a combination NC+SBC, with a significant contribution of blueshifted SBC components. BLUE are seen in all the sample and can indicate outflowing gas as suggested by many authors \citep{zakamska_2016, marziani_2016, canodiaz_2012}. In general, [O III]$\lambda\lambda$4959,5007 is wider in Pop. A than in Pop. B sources in our sample.

\par Regarding the UV region, we analyse CIV$\lambda$1549+HeII$\lambda$1640. In Fig. \ref{fig:PopAeB}, we show the fittings for one Pop. A and one Pop. B. In all the cases, the BC of CIV$\lambda$1549 is set at rest-frame and there is a strong contribution of blueshifted components to the full profile of CIV$\lambda$1549. The strongest BLUE components are seen in Pop. A and the redshifted VBC represent a significant part of the full profile of C IV$\lambda$1549 in Pop. B quasars. In the case of SDSSJ32012.33+142037.1 (the Pop. A one), it was necessary to include a second blueshifted component in order to fully reproduce the C IV$\lambda$1549 full profile. A similar behaviour is also seen in He II$\lambda$1640.

\begin{figure*}
    \centering
    \includegraphics[width=0.337\linewidth]{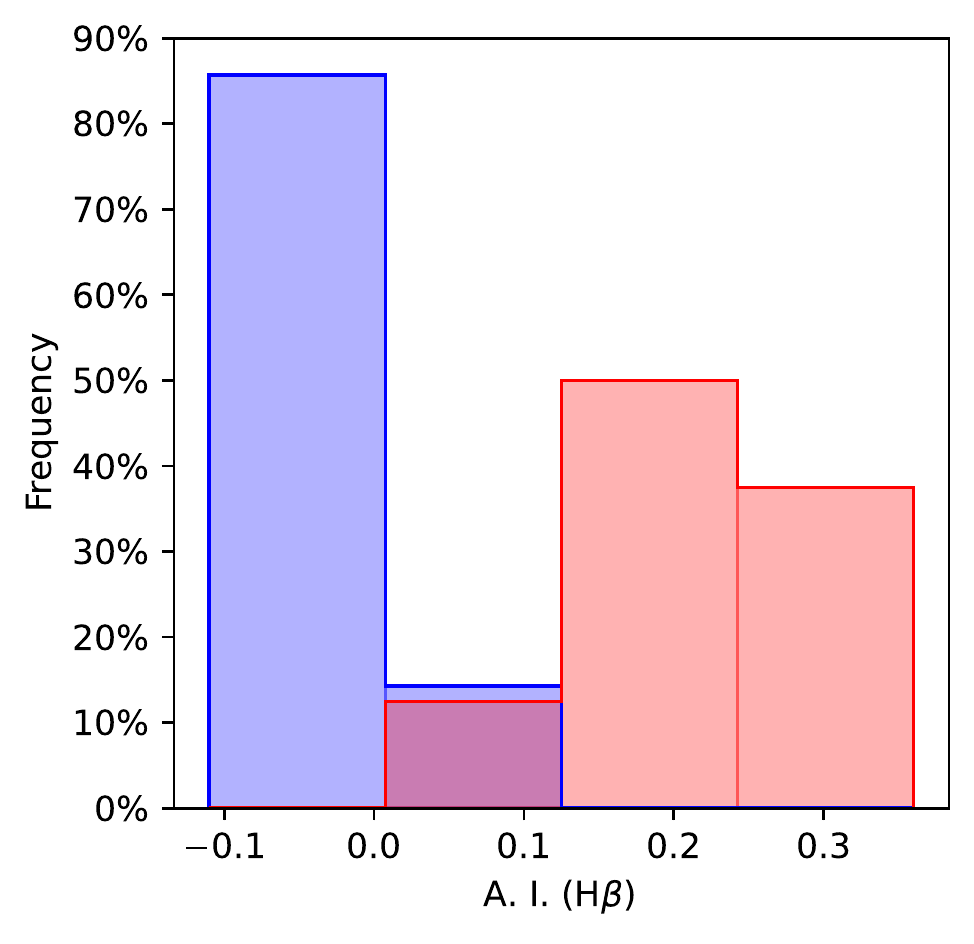}
    \includegraphics[width=0.32\linewidth]{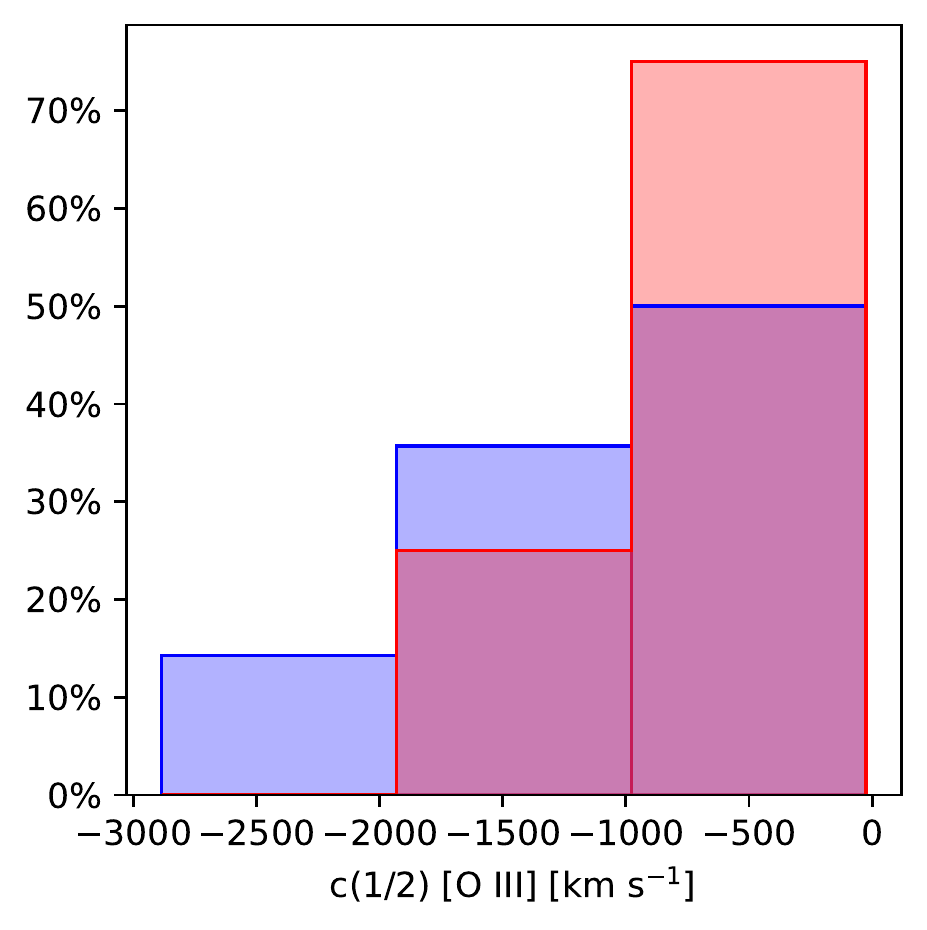}
    \includegraphics[width=0.32\linewidth]{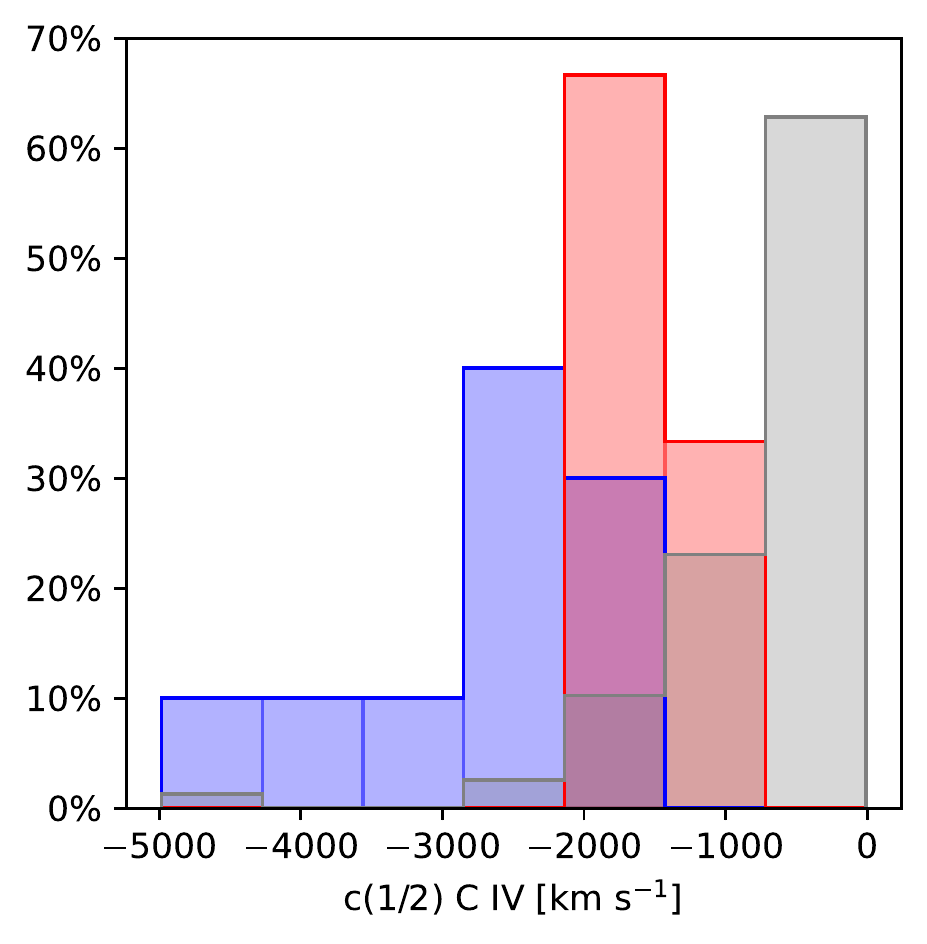}
    \caption{\textit{From left to right:} Histograms of H$\beta$ asymmetry, velocity centroid at 1/2 intensity (c(1/2)) of the [O III]$\lambda\lambda4959,5007$ emission line, and c(1/2) of C IV$\lambda$1549. Blue and red colors indicate Pop. A and Pop. B, respectively. Grey sample represents the data from \cite{sulentic_2007}.}
    \label{fig:histograms}
\end{figure*}

\begin{figure}[t]
    \centering
    \includegraphics[width=\columnwidth]{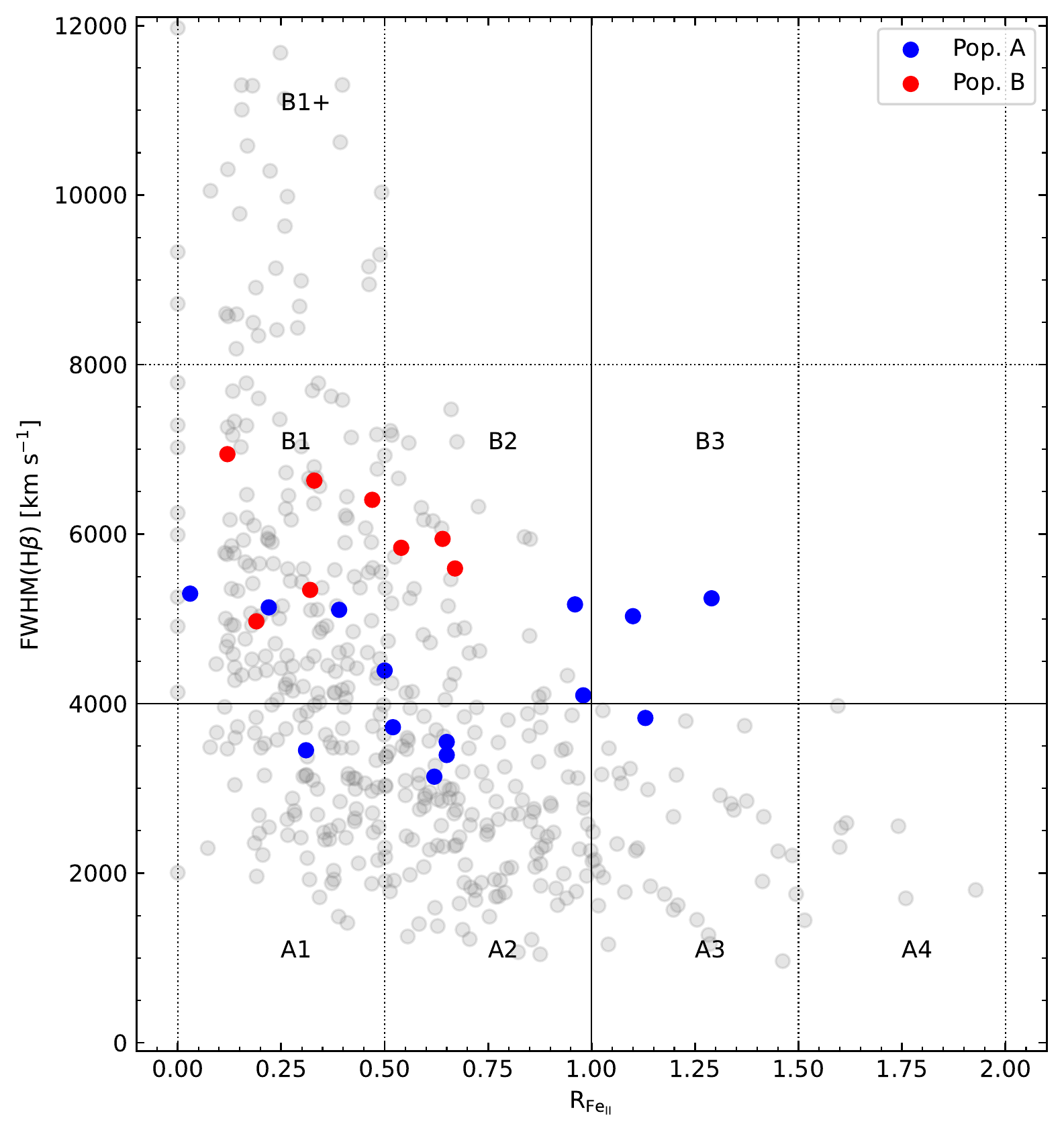}
    \caption{Location of the sample in the optical plane of the 4DE1. Pop. A are shown in blue and Pop. B in red. Filled circles represent the radio-quiet sources, while open circles show the radio-loud quasars. Grey circles are data from \cite{Zamfir_2008}.}
    \label{fig:optical_plane}
\end{figure}

\subsection{Full profiles}

\par Table \ref{tab:my_label} shows the source identification (Col. 1), the population classification (Col. 2), and the FWHM of the full profiles of H$\beta$ and C IV$\lambda$1549 (Cols. 3 and 4, respectively). We have found 14 out of the 22 sources to be Pop. A and 8 to be Pop. B. Since we could not find UV spectra with good S/N and that include C IV$\lambda$1549 for the complete sample, we analyse this emission line for only 13/22 quasars.
\par  In a comparison with Pop. A, Pop. B quasars are the ones that present a wider full profile of H$\beta$ with an FWHM average of $\sim$ 5959 km s$^{-1}$. For Pop. A sources this value is $\sim 4000$ km s$^{-1}$. On the other hand, the broadest C IV$\lambda$1549 full profiles are seen in Pop. A sources, especially in the ones classified as spectral type A3. On average, FWHM(C IV$_{\rm{full}}$) is $\sim$ 6830 km s$^{-1}$ for Pop. A and $\sim$ 5938 km s$^{-1}$ for Pop. B quasars, with an average FWHM ratio between CIV and H$\beta$ FWHM(CIV)/FWHM(H$\beta$)$=1.49$.
\par Fig. \ref{fig:histograms} shows histograms of the asymmetry of the H$\beta$ emission line, and velocity centroid at 1/2 intensity (c(1/2)) for [O III]$\lambda\lambda$4959,5007 and C IV$\lambda$1549. Pop. A and Pop. B are represented by blue and red colours, respectively. Considering first the histogram of H$\beta$ asymmetry, it is clear the difference between the behaviour of Pop. A and Pop. B sources. While Pop. B quasars tend to present a full profile shifted towards higher wavelengths, Pop. A H$\beta$ full profiles are centered at the rest-frame and are blueshifted. Only 5 out of the 14 Pop. A quasars present a significant outflow component.
\par Centre panel of Fig. \ref{fig:histograms} indicates the velocity centroid at half flux intensity (c(1/2)) for [O III]$\lambda\lambda$4959,5007 emission lines. In the two populations A and B, the majority of the sources present a small shift to the blue, in a bin going from -1000 km s$^{-1}$ to $\sim$ 0 km s$^{-1}$. The sources that present [O III]$\lambda\lambda$4959,5007 blueshifts at half flux intensity larger than 2000 km s$^{-1}$ are Pop. A quasars.
\par Similar behaviour happens in the C IV$\lambda$1549 profile, shown in the left histogram of Fig. \ref{fig:histograms}. The sources that present larger c(1/2) for C IV$\lambda$1549 are the Pop. A3 sources due to the presence of strong blueshifted components.  Grey data on the right plot of Fig. \ref{fig:histograms} are from \cite{sulentic_2007}, which is a sample with 70 low-redshift and low-luminosity RQ sources. In this case, the shifts at half intensity are very low when compared with our sample, with an averaged maximum of $\ge$ 1000 km s$^{-1}$.

\subsection{The location in the optical plane}
\par Fig. \ref{fig:optical_plane} shows the location of the sample in the 4DE1 optical plane. Blue and red points indicate Pop. A and Pop. B, respectively. Grey data are from the sample of \cite{Zamfir_2008} and characterize the Main Sequence of quasars at low redshift. As can be seen, our data present a displacement towards higher values of FWHM(H$\beta$) when compared to the sample of \cite{Zamfir_2008}. From Table \ref{tab:my_label}, we can see that there are some sources that present a FWHM(H$\beta_{\rm{full}}$) larger than 5000 km s$^{-1}$ as SDSSJ210524.47+000407.3 for instance. This is an expected behaviour in a high-luminosity context and the boundary FWHM(H$\beta$) $=$ 4000 km s$^{-1}$ should be a proxy used for lower redshifts \citep{marziani_2009, sulentic_2009}.


\section{Conclusions}

\par We analysed a sample of 22 high-redshift and high-luminosity QSOs, and found 14 of them to be Pop. A and 8 to be Pop. B quasars. The study is performed under the 4DE1 context, taking advantages of optical and UV properties from the sample. The main conclusions are:

\begin{itemize}
    \item Pop. B sources present broader full profiles of H$\beta$ than Pop. A, especially due to the virial VBC component shifted to the red;
    \item In general, C IV$\lambda$1549 emission line present very strong blueshifted components in the profile, indicating that winds and outflowing gas, or even jets may perform an important role in the quasar.
    \item There is a clear displacement to larger FWHM(H$\beta$) in the 4DE1 optical plane when considering high redshift sources like our sample.
\end{itemize}

\par However, a more detailed study is needed in order to perform a complete analysis to clarify the situation of high-redshift and high-luminosity sources in the complete context of the 4DE1.

\begin{acknowledgements} A. D. M. acknowledges the support of the INPhINIT fellowship from ``la Caixa'' Foundation (ID 100010434). The fellowship code is LCF/BQ/DI19/11730018. A. D. M. and A. d. O. acknowledge financial support form the State Agency for Research of the Spanish MCIU through the project PID2019-106027GB-C41 and the ``Center of Excellence Severo Ochoa'' award to the Instituto de Astrofísica de Andalucía (SEV-2017-0709). The authors acknowledge the SDSS for the optical spectra that were used in this work. \end{acknowledgements} 

\end{document}